\documentclass[12pt]{article}
\UseRawInputEncoding
\usepackage{amsfonts}
\usepackage{latexsym,amsmath}
\usepackage{amssymb,array}
\usepackage{setspace}
\usepackage[]{inputenc}

\makeatletter \oddsidemargin 0in \evensidemargin 0in \textwidth 16cm
 \RequirePackage[dvips]{graphicx} \textheight 18cm
\newcommand{\bi}{\begin{itemize}}
\newcommand{\ei}{\end{itemize}}
\newcommand{\bea}{\begin{eqnarray}}
\newcommand{\eea}{\end{eqnarray}}
\newtheorem{Definition}{Definition}[section]
\newtheorem{Theorem}{Theorem}[section]
\newtheorem{Lemma}{Lemma}[section]
\newcommand{\bl}{\begin{Lemma}}
\newcommand{\el}{\end{Lemma}}
\newtheorem{Proposition}{Proposition}[section]
\newtheorem{Notation}{Notation}[section]
\newcommand{\bn}{\begin{Notation}}
\newcommand{\en}{\end{Notation}}
\newtheorem{Remark}{Remark}[section]
\newcommand{\br}{\begin{Remark}}
\newcommand{\er}{\end{Remark}}
\newtheorem{Observation}[Definition]{Observation}
\newcommand{\bo}{\begin{Observation}}
\newcommand{\eo}{\end{Observation}}
\newtheorem{Construction}[Definition]{Construction}
\newcommand{\bcon}{\begin{Construction}}
\newcommand{\econ}{\end{Construction}}
\newcommand{\bex}{\begin{Example}}
\newcommand{\eex}{\end{Example}}
\newcommand{\bd}{\begin{Definition}}
\newcommand{\ed}{\end{Definition}}
\newcommand{\bs}{\begin{subsection}}
\newcommand{\es}{\end{subsection}}
\newtheorem{Corollary}{Corollary}[section]
\newtheorem{Example}{Example}[section]

\newcommand{\be}{\begin{equation}}
\newcommand{\ee}{\end{equation}}
\date{}
\begin{document}
    \newtheorem{r1}{Remark}[section]
    \newtheorem{co1}{Counterexample}[section]
    \newtheorem{note}{Note}[section]
    \numberwithin{equation}{section}
\title{Reliability study of a coherent system with single general standby component}
\author{Pradip Kundu$^1$, Nil Kamal Hazra$^2$ and Asok K. Nanda$^1$\footnote
    {Corresponding author, e-mail:
    asok.k.nanda@gmail.com}\\
    $^1$ Department of Mathematics and Statistics, IISER
Kolkata\\ Mohanpur 741246, India\\
$^2$ Department of Mathematical Statistics and Actuarial Science
\\ University of the Free State, 339 Bloemfontein 9300, South Africa}
\maketitle
\begin{abstract} The properties of a coherent system with a single general standby
component is investigated. Here three different switch over viz.
perfect switching, imperfect switching and random worm up period of
the standby component are considered with some numerical examples.
\end{abstract}
Keywords \& Phrases: General standby system, System signature.
\section{Introduction} Standby allocation is one of the widely used
techniques to improve the reliability of a system. Standby
components are mostly of three types - cold standby, warm standby
and hot standby. Cold standby means that the redundant component is
inactive and has zero failure rate while in standby, and it starts
to function at the time when the system/component fails. Hot standby
describes the scenario where the redundant component and the
corresponding system undergoes the same operational environment. In
case of warm standby, the redundant component undergoes two
operational environments, namely, usual environment (the environment
in which the system is running) and milder environment (where the
redundant component has less failure rate than that in the usual
environment). Initially the warm standby component functions in
milder environment and then it switches over to the usual
environment at the time of system/component failure. Papageorgiou
and Kokolakis \cite{papa} derived the reliability function of a two
component parallel system with $(n-2)$ warm standby components,
where two units start their operation simultaneously and any one of
them is replaced instantaneously upon its failure by one of the
$(n-2)$ warm standbys. Cha et al. \cite{cha} introduced a general
standby model for a single-component system with a single standby
component, and derived system performance measures. The cold and the
hot standby models are derived as special cases of the general
standby model. Li et al. \cite{xli1,xli} investigated some general
standby systems and derived some stochastic comparison results on
the lifetimes of the systems. Hazra and Nanda \cite{hazra} discussed
some standby models with one and two general standby components, and
compared some different series and parallel systems corresponding to
the models with respect to the usual stochastic and the stochastic
precedence orders. Eryilmaz \cite{eryl} derived reliability function
of a coherent system equipped with a cold standby component such
that the coherent system may fail at the time of the first component
failure. Recently Franko et al. \cite{franko} generalized this case
by considering that the coherent system may fail at the time of
$s$th component failure so that the standby component may be put
into operation actively at the time of $s$-th component failure. In
case of warm standby redundancy, Eryilmaz \cite{ery2} investigated
the reliability properties of $k$-out-of-$n$ system equipped with a
single warm standby component.\par In this paper, we investigate the
reliability properties of a coherent system equipped with a general
standby component. It is to be mentioned here that the general
standby studied in this paper has three different states. To be more
specific, the standby component starts to work in cold state, it is
switched over to warm state (from cold state) after a specified time
$u\; (\geq 0)$ and put into operation in active state in usual
environment at the time of $s$th component failure which might cause
the system to fail. Switching from cold state to warm state of the
standby component takes place after a specified time $u\; (>0)$ only
if it is known a priori that $s$th component failure does not occur
before time $u\; (>0)$; otherwise, $u$ must be zero, i.e., standby
component starts to work in warm state from the beginning. We also
consider the perfectness of the switching from one state to another
state of the standby component.\par The rest of the paper is
organized as follows. Section \ref{sec2} discusses briefly general
standby model and representation of survival function of a coherent
system using system signature. In Section \ref{sec3}, we obtain
reliability of a coherent system equipped with a single general
standby component in different switch over cases, namely perfect
switching case, imperfect switching case and the case with random
warm-up period. Numerical example is presented in this section.
Finally the paper is concluded in Section \ref{sec4}.\par For two
random variables $X$ and $Y$, $X=^{st}Y$ means that $X$ and $Y$ have
the same distribution.
\section{General standby model}\label{sec2} The concept
of accelerated life tests and that of virtual age (see also Kijima
\cite{kijima}, Finkelstein~ \cite{fink}) have been used by Cha et
al. \cite{cha} for modeling general standby system in case of a
single active component system. Let $X$ be the lifetime of an active
component with cumulative distribution function (c.d.f.) $F(\cdot)$
and let $Y$ be the lifetime of a standby component in the usual
environment with c.d.f. $G(\cdot)$. Further, let $X$ and $Y$ be
independent. Write $\bar{F}(\cdot)=1-F(\cdot)$ and
$\bar{G}(\cdot)=1-G(\cdot)$. For the standby unit in warm state, the
component operates in an environment which is milder than the usual
level of environment. Thus, the lifetime of the standby component in
warm state will have the c.d.f. $G(\gamma(\cdot))$ where
$\gamma(\cdot)$ is a non-decreasing function satisfying
$\gamma(t)\leq t$, for all $t\geq 0$ with $\gamma(0)=0$. Now,
suppose that the standby unit has worked during $(0,t]$ without
failure in a warm state, and is activated under usual environment at
time $t$. Then the virtual age $\omega(t)$ of the standby component
is non-decreasing satisfying $\omega(t)\leq t$, for all $t\geq 0$
and $\omega(0)=0$. Let $Y^\ast$ denote the remaining lifetime of the
standby component after the failure of the active unit at time
$X=x$. Then (cf. Cha et al. \cite{cha})
$$P\{Y^\ast>t, I=1\mid
X=x\}=\frac{\bar{G}(\omega(x)+t)}{\bar{G}(\omega(x))}
\bar{G}(\gamma(x))$$ and the survival function of the standby system
is \begin{equation}\label{eq1st}R(t)= \bar{F}(t)+\int_{0}^t
\frac{\bar{G}(\omega(x)+t-x)}{\bar{G}(\omega(x))} \bar{G}(\gamma(x))
dF(x),\end{equation} where $\{I=1\}$ indicates that the standby
component survives at least up to the failure time of the active
component. The cold and the hot standby models can be derived as
special cases by substituting $\gamma(t)=\omega(t)=0$ and
$\gamma(t)=\omega(t)=t$, respectively.
\par In the above discussion, the standby component starts to work in warm state
at the beginning. However, for a general standby component, it can
be assumed that the main (active) component starts to work in active
state, and the standby component starts to work in cold state, and
is switched over to warm state after a pre-specified time $u$ up to
which the active component certainly does not fail. In this case, if
the standby unit operates during $(u,x]$ without failure in warm
state, then obviously the virtual age at time $x$ would be
$\omega(x-u)$. Now to survive the system up to time $t$, the
following cases may arise. The active component survives up to time
$t$, or fails in $(u,t]$ and the standby component survives for the
remaining time. Clearly, if $t\leq u$, then the reliability of the
system is $\bar{F}(t)$; otherwise (cf. Yun and Cha \cite{yuncha}),
\begin{equation}\label{eq1gst}R(t)= \bar{F}(t)+\int_{u}^t
\frac{\bar{G}(\omega(x-u)+t-x)}{\bar{G}(\omega(x-u))}
\bar{G}(\gamma(x-u)) dF(x).\end{equation}
\section{Main results}\label{sec3}
Let $T$ denote the lifetime of a binary coherent system without a
standby and let $T^{gs}$ denote the lifetime of the same system with
a general standby component whose lifetime is $Y$. Suppose $F$ is
the common absolutely continuous c.d.f. of
$X_{1},X_{2},\ldots,X_{n}$ having probability density function
(p.d.f.) $f$, and $G$ is the absolutely continuous c.d.f. of $Y$
having p.d.f. $g$, where $X_{i}$ denoteS the lifetime of the $i$th
component, $i=1,2,\ldots,n$. Here, $Y$ and
$X_{1},X_{2},\ldots,X_{n}$ are independent. Eryilmaz \cite{ery2}
studied the reliability properties of the $k$-out-of-$n$ system with
a single warm standby component. Franko et al. \cite{franko} studied
coherent systems equipped with a cold standby component which may be
put into operation at the time of the $s$th component failure,
$s=k_{\phi},k_{\phi}+1,\ldots,z_{\phi}+1$, where $k_{\phi}$ is the
minimum number of failed components that causes the system failure
whereas $z_{\phi}$ is the maximum number of failed components with
which the system can still operate. In this paper, we consider
coherent system equipped with a general standby component which may
start to work in cold state, and is switched over to warm state
after a specified time $u(\geq 0)$; after that it may be put to work
actively in the usual environment at the time of the $s$th component
failure which may cause the system failure. It is clear that the
standby component operated under warm standby state may be put to
work in the usual environment if the system has a positive
probability of failure at the time of the $s$th component failure.
This means that $P\{T=X_{s:n}\}>0$, for
$s=k_{\phi},k_{\phi}+1,\ldots,z_{\phi}+1$.\par Now, after putting
the warm standby component into operation in the usual environment
at the time when the system fails upon the failure of the $s$th
component, the remaining lifetime of the system consisting of
$(s-1)$ failed components (0's), a standby component, and $(n-s)$
working components, can be represented as
$$\phi_{s}(0_{B_{1}},0_{B_{2}},\ldots,0_{B_{s-1}},Y_{V_{s}}^{(u,s)},
X_{R_{1}}^{(s)},X_{R_{2}}^{(s)},\ldots,X_{R_{n-s}}^{(s)}),$$ where
$V_{s}$ is the discrete random variable representing the index of
the standby component indicating that it is put into operation from
warm state to active state when $s$th component fails,
$Y_{V_{s}}^{(u,s)}$ is the remaining lifetime of the general standby
component after failure of the $s$th component (where $u$ indicates
the time after which the standby component starts to work in warm
state), $B_{i}\in \{1,2,\ldots,n\}$, $i=1,2,\ldots,s-1$ and
$R_{j}\in \{1,2,\ldots,n\}$, $j=1,2,\ldots,n-s$ are the discrete
random variables representing respectively the index of the $i$th
failed component and the $j$th surviving component at the time of
$s$th component failure, and consequently $X_{l}^{(s)}$ represents
the remaining lifetime of the $l$th component after failure of the
$s$th component, i.e., $X_{l}^{(s)}=^{st}(X_{l}-X_{s:n}\mid
X_{l}>X_{s:n})$, $1\leq l\leq n-s$, $X_{s:n}$ is the $s$th order
statistic from $X_{1},X_{2},\ldots,X_{n}$. It is to be noted that
$V_{s}=c$ if and only if $(X_{c}=X_{s:n}\mid
T=X_{s:n}),c=1,2,\ldots,n$. The reliability properties of the
remaining lifetime of the system is computed based on $(n-s+1)$
functioning components (including the standby component). Note that,
here only the places of the $(s-1)$ failed components are taken into
consideration in the structure function, and not their lifetimes (as
they have already failed).\par Our main goal is to investigate the
reliability characteristics of $T^{gs}$ which is the lifetime of the
system with general standby component, i.e.,
$$T^{gs}=T+\sum_{s=k_{\phi}}^{z_{\phi}+1}
\phi_{s}(0_{B_{1}},0_{B_{2}},\ldots,0_{B_{s-1}},Y_{V_{s}}^{(u,s)},X_{R_{1}}^{(s)},X_{R_{2}}^{(s)},\ldots,X_{R_{n-s}}^{(s)}).$$
The following proposition will be used to calculate the survival
functions of the system with a general standby.
\begin{Proposition}\label{prop1}For $t\geq
x\geq u\geq 0$ and $s=k_{\phi},k_{\phi}+1,\ldots,z_{\phi}+1$,
\begin{eqnarray*}\nonumber && P\{\phi_{s}(0_{b_{1}},0_{b_{2}},\ldots,0_{b_{s-1}},Y_{c}^{(u,s)},X_{r_{1}}^{(s)},
X_{r_{2}}^{(s)},\ldots,X_{r_{n-s}}^{(s)})>t\mid X_{s:n}=x\}
\\ \nonumber
&&=\frac{\bar{G}(\gamma(x-u))}{\bar{G}(\omega(x-u))\bar{F}^{n-s}(x)}
\idotsint\limits_{\phi_{s}(0_{b_{1}},0_{b_{2}},\ldots,0_{b_{s-1}},y+u,x_{r_{1}},x_{r_{2}},\ldots,x_{r_{n-s}})>t}
g(\omega(x-u)+y+u)\\&& \prod_{m=1}^{n-s} f(x_{r_{m}}+x)
dx_{r_{1}}dx_{r_{2}}\ldots dx_{r_{n-s}}dy.
\end{eqnarray*}
\end{Proposition}
\textbf{Proof:} Because $Y$ and $X_{1},X_{2},\ldots,X_{n}$ are
independent,
\begin{eqnarray*}&&
P\{\phi_{s}(0_{b_{1}},0_{b_{2}},\ldots,0_{b_{s-1}},Y_{c}^{(u,s)},X_{r_{1}}^{(s)},X_{r_{2}}^{(s)},\ldots,X_{r_{n-s}}^{(s)})>t\mid
X_{s:n}=x\}\\&&=\idotsint\limits_{\phi_{s}(0_{b_{1}},0_{b_{2}},\ldots,0_{b_{s-1}},y+u,x_{r_{1}},x_{r_{2}},\ldots,x_{r_{n-s}})>t}
g(y+u\mid
X_{s:n}=x)\times\\&&f_{J}(x_{r_{1}},x_{r_{2}},\ldots,x_{r_{n-s}}\mid
X_{s:n}=x) dx_{r_{1}}dx_{r_{2}}\ldots dx_{r_{n-s}}dy,
\end{eqnarray*} where $f_{J}$ denotes the joint p.d.f. of
$X_{r_{1}}^{(s)},X_{r_{2}}^{(s)},\ldots,X_{r_{n-s}}^{(s)}$ given that
$X_{s:n}=x$. Note that
\begin{equation}\nonumber f_{J}(x_{r_{1}},x_{r_{2}},\ldots,x_{r_{n-s}}\mid
X_{s:n}=x)=\frac{1}{\bar{F}^{n-s}(x)}\prod_{m=1}^{n-s}
f(x_{r_{m}}+x),\end{equation} since the random variables
$X_{r_{1}}^{(s)},X_{r_{2}}^{(s)},\ldots,X_{r_{n-s}}^{(s)}$ are
conditionally independent given that $X_{s:n}=x$, and
$$P\{X_{r_{1}}^{(s)}>x_{r_{1}},X_{r_{2}}^{(s)}>x_{r_{2}},\ldots,X_{r_{n-s}}^{(s)}>x_{r_{n-s}}\mid
X_{s:n}=x\}=\prod_{m=1}^{n-s}\frac{\bar{F}(x_{r_{m}}+x)}{\bar{F}(x)}.$$
Also
\begin{eqnarray}\nonumber G(y+u\mid
X_{s:n}=x)&=&1-P\{Y_{c}^{(u,s)}>y+u\mid X_{s:n}=x\}\\\nonumber
&=&1-\frac{\bar{G}(\omega(x-u)+y+u)}{\bar{G}(\omega(x-u))}\bar{G}(\gamma(x-u)).
\end{eqnarray}
Now the result follows after simplification.$\hfill\Box$
\subsection{Reliability of a coherent system with perfect
switching}\label{sec2.1} Consider a coherent system having $n$
components and one general standby component. Initially, $n$
components start working in active state and the standby component
starts to work in cold state. As we have mentioned earlier, the
standby component is switched over to the warm state (from the cold
state) after a specified time $u\;(\geq 0)$ up to which the system
certainly does not fail, and it starts to work in active state in
the usual environment at the time of $s$th component failure which
may cause the system failure. Obviously, for $u=0$, the system
becomes a coherent system equipped with a warm standby component.
Here we assume that, for the standby component, switching from cold
state to warm state and from warm state to active state are perfect,
i.e., the standby component does not fail at the time of switch over
from one state to another, and that it is instantaneous. The
reliability of such a system is given by the following theorem.
Before going to the theorem, we briefly discuss the representation
of survival function of a coherent system using system signature.
Let $T$ denote the lifetime of a coherent system consisting of $n$
components having lifetimes $X_{1},X_{2},\ldots,X_{n}$. If $X_{i}$'s
are independent and have common distribution, then the survival
function can be given by
$$P\{T>t\}=\sum_{i=1}^n p_{i} P\{X_{i:n}>t\},$$ where
$p_{i}=P\{T=X_{i:n}\}$ is the probability that the $i$th component
failure causes the system to fail. The $n$-dimensional probability
vector $\textbf{p}=(p_{1},p_{2},\ldots,p_{n})$ is called system
signature (cf. Kocher et al. \cite{kochar}, Samaniego
(\cite{saman1},\cite{saman2})).\par Let $\textbf{B}_{s,c}$ represent
the discrete multivariate random variable representing the indices
of the failed component given that $V_{s}=c$; $s,c\in
\{1,2,\ldots,n\}$, i.e.,
$\textbf{B}_{s,c}=(B_{1}=b_{1},B_{2}=b_{2},\ldots
B_{s-1}=b_{s-1}\mid V_{s}=c)$ if and only if
$(0_{B_{1}}=0_{b_{1}},0_{B_{2}}=0_{b_{2}},\ldots,0_{B_{s-1}}=0_{b_{s-1}}\mid
X_{c}=X_{s:n}, T=X_{s:n}),$ where $B_{i}\in \{1,2,\ldots,n\}$,
$i=1,2,\ldots,s-1$, is the discrete random variable representing the
index of the $i$th component failure. The following theorem gives
the reliability function of a coherent system with a general standby
component in case of perfect switching. The calculation has been
illustrated in Example \ref{aaa}.
\begin{Theorem}\label{th1}Let $\textbf{p}$ be the signature vector of
a coherent system which is equipped with a general standby
component. Then, for $t>u$,
{\small\begin{eqnarray*}  P\{T^{gs}>t\}&=&\sum_{s=k_{\phi}}^{z_{\phi}+1}\big(p_{s}P\{X_{s:n}>t\}+p_{s}\sum_{c=1}^nP\{V_{s}=c\}\sum_{1\leq b_{1}<\ldots<b_{s-1}\leq n}P\{B_{s,c}=(b_{1},\ldots,b_{s-1})\})\\
&& \times \int_{u}^t P\{\phi_{s}(0_{b_{1}},0_{b_{2}},\ldots,0_{b_{s-1}},Y_{c}^{(u,s)},X_{r_{1}}^{(s)},X_{r_{2}}^{(s)},\ldots,X_{r_{n-s}}^{(s)})>t-x\mid X_{s:n}=x\}dF_{s:n}(x)\big).
\end{eqnarray*}}
\end{Theorem}
\textbf{Proof:} To derive the system reliability the following cases
may arise in which the system survives at time $t>u$. Under the
condition $P\{T=X_{s:n}\}>0$, for
$s=k_{\phi},k_{\phi}+1,\ldots,z_{\phi}+1$, any coherent system
operating with $n$ active components may fail at the time of $s$th
component failure. If the system failure occurs due to the $s$th
component failure in $(u,t]$, then the warm standby component is
switched over to usual environment. Thus, the survival function of
the system can be written as

{\scriptsize \allowdisplaybreaks{
\begin{eqnarray}\nonumber P\{T^{gs}>t\}&=&P\{T+\phi_{k_{\phi}}(0_{B_{1}},0_{B_{2}},\ldots,0_{B_{k_{\phi}-1}},Y_{V_{k_{\phi}}}^{(u,k_{\phi})},X_{R_{1}}^{(k_{\phi})},X_{R_{2}}^{(k_{\phi})},\ldots,X_{R_{n-k_{\phi}}}^{(k_{\phi})})>t,
T=X_{k_{\phi}:n}\}\\\nonumber&&+P\{T+\phi_{k_{\phi}+1}(0_{B_{1}},0_{B_{2}},\ldots,0_{B_{k_{\phi}}},Y_{V_{k_{\phi}+1}}^{(u,k_{\phi}+1)},X_{R_{1}}^{(k_{\phi}+1)},X_{R_{2}}^{(k_{\phi}+1)},\ldots,X_{R_{n-k_{\phi}-1}}^{(k_{\phi}+1)})>t,
T=X_{k_{\phi}+1:n}\}+\ldots\\\nonumber&&+P\{T+\phi_{z_{\phi}+1}(0_{B_{1}},0_{B_{2}},\ldots,0_{B_{z_{\phi}}},Y_{V_{z_{\phi}+1}}^{(u,z_{\phi}+1)},X_{R_{1}}^{(z_{\phi}+1)},X_{R_{2}}^{(z_{\phi}+1)},\ldots,X_{R_{n-z_{\phi}-1}}^{(z_{\phi}+1)})>t,
T=X_{z_{\phi}+1:n}\}\\\nonumber&&+P\{T>t,T>X_{z_{\phi}+1:n}\}.\\\nonumber&=&p_{k_{\phi}}P\{T+\phi_{k_{\phi}}(0_{B_{1}},0_{B_{2}},\ldots,0_{B_{k_{\phi}-1}},Y_{V_{k_{\phi}}}^{(u,k_{\phi})},X_{R_{1}}^{(k_{\phi})},X_{R_{2}}^{(k_{\phi})},\ldots,X_{R_{n-k_{\phi}}}^{(k_{\phi})})>t\mid
T=X_{k_{\phi}:n}\}+p_{k_{\phi}+1}\\\nonumber&&P\{T+\phi_{k_{\phi}+1}(0_{B_{1}},0_{B_{2}},\ldots,0_{B_{k_{\phi}}},Y_{V_{k_{\phi}+1}}^{(u,k_{\phi}+1)},X_{R_{1}}^{(k_{\phi}+1)},X_{R_{2}}^{(k_{\phi}+1)},\ldots,X_{R_{n-k_{\phi}-1}}^{(k_{\phi}+1)})>t\mid
T=X_{k_{\phi}+1:n}\}+\ldots\\\nonumber
&&+p_{z_{\phi}+1}P\{T+\phi_{z_{\phi}+1}(0_{B_{1}},0_{B_{2}},\ldots,0_{B_{z_{\phi}}},Y_{V_{z_{\phi}+1}}^{(u,z_{\phi}+1)},X_{R_{1}}^{(z_{\phi}+1)},X_{R_{2}}^{(z_{\phi}+1)},\ldots,X_{R_{n-z_{\phi}-1}}^{(z_{\phi}+1)})>t\mid
T=X_{z_{\phi}+1:n}\},\label{eqth-1} \\
\end{eqnarray}}}
since
$P\{T>t,T>X_{z_{\phi}+1:n}\}=0$. Now, write $\textbf{0}_{B}$ as
the event that $(0_{B_{1}}=0_{b_{1}},\ldots,0_{B_{s-1}}=0_{b_{s-1}})$.
Then, for $s=k_{\phi},k_{\phi}+1,\ldots ,z_{\phi}+1$ and $t>u(\geq 0)$, we have
 \allowdisplaybreaks{
\begin{eqnarray*}&&P\{T+\phi_{s}(0_{B_{1}},0_{B_{2}},\ldots,0_{B_{s-1}},Y_{V_{s}}^{(u,s)},X_{R_{1}}^{(s)},X_{R_{2}}^{(s)},\ldots,X_{R_{n-s}}^{(s)})>t\mid
T=X_{s:n}\}\\
&=&\sum_{c=1}^n\frac{P\{X_{c}+\phi_{s}(0_{B_{1}},0_{B_{2}},\ldots,0_{B_{s-1}},Y_{c}^{(u,s)},X_{R_{1}}^{(s)},X_{R_{2}}^{(s)},\ldots,X_{R_{n-s}}^{(s)})>t,X_{s:n}=X_{c},T=X_{s:n}\}}{P\{T=X_{s:n}\}}\\ &=&\sum_{c=1}^nP\{X_{c}+\phi_{s}(0_{B_{1}},0_{B_{2}},\ldots,0_{B_{s-1}},Y_{c}^{(u,s)},X_{R_{1}}^{(s)},X_{R_{2}}^{(s)},\ldots,X_{R_{n-s}}^{(s)})>t\mid
X_{s:n}=X_{c},T=X_{s:n}\}\\&& \times P\{X_{s:n}=X_{c}\mid
T=X_{s:n}\}\\
 &=&\left(P\{X_{s:n}=X_{c},T=X_{s:n}\}\right)^{-1} \left[\sum_{c=1}^nP\{V_{s}=c\}
\sum\limits_{1\leq
b_{1}<\ldots<b_{s-1}\leq n}
P\{\textbf{0}_{B},X_{s:n}=X_{c}, T=X_{s:n}\}\times\right.\\&&\left. P\{X_{c}+\phi_{s}(0_{b_{1}},0_{b_{2}},\ldots,0_{b_{s-1}},Y_{c}^{(u,s)},X_{r_{1}}^{(s)},X_{r_{2}}^{(s)},\ldots,X_{r_{n-s}}^{(s)})>t\mid
\textbf{0}_{B}, X_{s:n}=X_{c}, T=X_{s:n}\}\right]\\
&=&\sum_{c=1}^nP\{V_{s}=c\}\sum\limits_{1\leq b_{1}<\ldots<b_{s-1}\leq
n}P\{\textbf{0}_{B}\mid X_{s:n}=X_{c}, T=X_{s:n}\}\\&& \times
P\{X_{c}+\phi_{s}(0_{b_{1}},0_{b_{2}},\ldots,0_{b_{s-1}},Y_{c}^{(u,s)},X_{r_{1}}^{(s)},X_{r_{2}}^{(s)},\ldots,X_{r_{n-s}}^{(s)})>t\mid\textbf{0}_{B},
X_{s:n}=X_{c}, T=X_{s:n}\}\\
&=&\sum_{c=1}^nP\{V_{s}=c\}\sum\limits_{1\leq b_{1}<\ldots<b_{s-1}\leq
n}P\{\textbf{B}_{s,c}=(b_{1},\ldots,b_{s-1})\} \\&& \times \int
P\{\phi_{s}(0_{b_{1}},0_{b_{2}},\ldots,0_{b_{s-1}},Y_{c}^{(u,s)},X_{r_{1}}^{(s)},X_{r_{2}}^{(s)},\ldots,X_{r_{n-s}}^{(s)})>t-x\mid
X_{s:n}=x\}dF_{s:n}(x)\\
&=&\sum_{c=1}^nP\{V_{s}=c\}\sum\limits_{1\leq b_{1}<\ldots<b_{s-1}\leq
n}P\{\textbf{B}_{s,c}=(b_{1},\ldots,b_{s-1})\}\times
\left[\int_{t}^{\infty}dF_{s:n}(x)+\right.
\\&& \left. \int_{u}^t
P\{\phi_{s}(0_{b_{1}},0_{b_{2}},\ldots,0_{b_{s-1}},Y_{c}^{(u,s)},X_{r_{1}}^{(s)},X_{r_{2}}^{(s)},\ldots,X_{r_{n-s}}^{(s)})>t-x\mid
X_{s:n}=x\}dF_{s:n}(x)\right]\end{eqnarray*}}
\begin{equation}\nonumber \hspace{-3.5cm}
=P\{X_{s:n}>t\}+\sum_{c=1}^nP\{V_{s}=c\}\sum_{1\leq
b_{1}<\ldots<b_{s-1}\leq
n}P\{B_{s,c}=(b_{1},\ldots,b_{s-1})\}
\end{equation}
\begin{equation}\label{eqth2} \times \int_{u}^t
P\{\phi_{s}(0_{b_{1}},0_{b_{2}},\ldots,0_{b_{s-1}},Y_{c}^{(u,s)},X_{r_{1}}^{(s)},X_{r_{2}}^{(s)},\ldots,X_{r_{n-s}}^{(s)})>t-x\mid
X_{s:n}=x\}dF_{s:n}(x).
\end{equation} Now using (\ref{eqth2}), from
(\ref{eqth-1}) we have the required result.$\hfill\Box$\\
Below we give the result corresponding to the coherent system with
single warm standby component.
\begin{Corollary} If the standby
component starts to work in the warm state at time zero, then we get
the reliability
function of a coherent system equipped with single warm standby component from Theorem \ref{th1} by taking $u=0$.$\hfill\Box$
\end{Corollary}
\begin{Corollary} \label{coro}
 Reliability function of a $k$-out-of-$n$
system equipped with a general standby component is given by
\begin{eqnarray*}P\{T^{gs}>t\}
&=&P\{X_{n-k+1:n}>t\}+\frac{\bar{F}^{k-1}(t)}{B(n-k+1,k)}\\&& \times
\int_{u}^t \frac{\bar{G}(\omega(x-u)+t-x)}{\bar{G}(\omega(x-u))}
\bar{G}(\gamma(x-u))F^{n-k}(x)dF(x).
\end{eqnarray*}
\end{Corollary}
\textbf{Proof:}
For a $k$-out-of-$n$ system, we have $s=k_{\phi}=z_{\phi}+1=n-k+1$.
Because the signature of a $k$-out-of-$n$ system is the
$n$-dimensional vector $\textbf{p}=(0,\ldots,0,1,0,\ldots,0)$ with 1
is in the $(n-k+1)$th place, we have, from Theorem \ref{th1},
\begin{eqnarray*}
&&P\{T^{gs}>t\}=P\{X_{n-k+1:n}>t\}+\\ && \int_{u}^t
P\{\phi_{s}(0_{b_{1}},0_{b_{2}},\ldots,0_{b_{n-k}},Y_{n-k+1}^{(u,s)},X_{r_{1}}^{(s)},X_{r_{2}}^{(s)},\ldots,X_{r_{k-1}}^{(s)})>t-x\mid
X_{n-k+1:n}=x\}dF_{n-k+1:n}(x),\end{eqnarray*} with $s=n-k+1$. Now,
from Proposition \ref{prop1}, we have
\begin{eqnarray*}&&P\{\phi_{s}(0_{b_{1}},0_{b_{2}},\ldots,0_{b_{n-k}},Y_{n-k+1}^{(u,s)},X_{r_{1}}^{(s)},X_{r_{2}}^{(s)},\ldots,X_{r_{k-1}}^{(s)})>t-x\mid
X_{n-k+1:n}=x\}\\&=&\frac{\bar{G}(\gamma(x-u))}{\bar{G}(\omega(x-u))\bar{F}^{k-1}(x)}\times\\&&
\idotsint\limits_{\min(y+u,x_{r_{1}},x_{r_{2}},\ldots,x_{r_{k-1}})>t-x}
g(\omega(x-u)+y+u)\prod_{m=1}^{k-1} f(x_{r_{m}}+x) %\\&&
dx_{r_{1}}dx_{r_{2}}\ldots dx_{r_{k-1}}dy
\\&=&\frac{\bar{G}(\gamma(x-u))}{\bar{G}(\omega(x-u))\bar{F}^{k-1}(x)}
\bar{G}(\omega(x-u)+t-x)\bar{F}^{k-1}(t),\end{eqnarray*} so that
\begin{eqnarray}\nonumber P\{T^{gs}>t\}&=&P\{X_{n-k+1:n}>t\}+\\&& \nonumber \bar{F}^{k-1}(t)\int_{u}^t \frac{\bar{G}(\omega(x-u)+t-x)}{\bar{G}(\omega(x-u))}
\bar{G}(\gamma(x-u))\frac{1}{\bar{F}^{k-1}(x)}dF_{n-k+1:n}(x)\\\nonumber
&=&P\{X_{n-k+1:n}>t\}+\frac{\bar{F}^{k-1}(t)}{B(n-k+1,k)}\\&&
\label{eqk/n} \int_{u}^t
\frac{\bar{G}(\omega(x-u)+t-x)}{\bar{G}(\omega(x-u))}
\bar{G}(\gamma(x-u))F^{n-k}(x)dF(x),
\end{eqnarray}
where
$B(a,b)=\frac{\Gamma a \Gamma b}{\Gamma(a+b)}$. Now, for $u=0$,
(\ref{eqk/n}) gives the reliability function of a $k$-out-of-$n$
system equipped with a
warm standby component as shown in Eryilmaz \cite{ery2}.
\begin{Remark}\normalfont
Taking $n=k=1$ in Corollary \ref{coro}, we get the reliability of a
single (active) component system with a general standby unit as
given in (\ref{eq1gst}). In particular, for $u=0$, we obtain
(\ref{eq1st}).
\end{Remark}
\begin{Remark}\normalfont
By taking $u=0$ and $\gamma(x)=\omega(x)=0$ in Theorem \ref{th1}, we
obtain the reliability function of a coherent system equipped with a
cold standby component given in Franko et al.
\cite{franko}.$\hfill\Box$
\end{Remark}
The result of Theorem \ref{th1} can also be represented as follows.
\begin{Theorem}\label{th2}Let $\textbf{p}$ be the signature vector of a
coherent system equipped with a general standby component with
lifetime distribution $G$. Then {\footnotesize
\begin{eqnarray*}P\{T^{gs}>t\}&=&\sum_{s=k_{\phi}}^{z_{\phi}+1}\big(p_{s}P\{X_{s:n}>t\}+p_{s}\sum_{c=1}^nP\{V_{s}=c\}\sum_{1\leq
b_{1}<\ldots<b_{s-1}\leq n}P\{B_{s,c}=(b_{1},\ldots,b_{s-1})\}
\\&&\times \int_{u}^t
\frac{\bar{G}(\omega(x-u)+t-x)}{\bar{G}(\omega(x-u))}
\bar{G}(\gamma(x-u))\left[\sum_{k=1}^{n-s}p_{k}^{c,(b_{1},\ldots,b_{s-1})}P\{X_{k:n-s}^{(s)}>t-x\mid
X_{s:n}=x\}\right.\\&&\left.
+p_{n-s+1}^{c,(b_{1},\ldots,b_{s-1})}\right]dF_{s:n}(x)\big),\end{eqnarray*}}
where $\{p_{k}^{c,(b_{1},\ldots,b_{s-1})},~k=1,2,\ldots,n-s+1\}$
represents the signature vector corresponding to the $(n-s)$
surviving components and the standby component at place $c$ with the
understanding that when $p_{k}^{c,(b_{1},\ldots,b_{s-1})}$ is
non-zero for at least one $k\in\{1,2,\ldots,n-s\}$, then
$p_{n-s+1}^{c,(b_{1},\ldots,b_{s-1})}=0$.
\end{Theorem}
\textbf{Proof:}
Let the standby component switch over to the usual environment from
the warm state at the time $x\in [u,t]$, for $u\geq 0$. Then the
system may fail due to the failure of any of the $(n-s)$ surviving
components (other than the standby) or due to the failure of the
standby component. Clearly, if $p_{n-s+1}^{c,(b_{1},\ldots,b_{s-1})}$
is the probability that the system fails due to the failure of the
standby component, then $p_{n-s+1}^{c,(b_{1},\ldots,b_{s-1})}$ will be
zero if and only if at least one of $p_{k}^{c,(b_{1},\ldots,b_{s-1})}$,
$k\in\{1,2,\ldots,n-s\}$ is non-zero, otherwise
$p_{n-s+1}^{c,(b_{1},\ldots,b_{s-1})}$ will be unity. Then, we have
\begin{eqnarray*}&&
P\{\phi_{s}(0_{b_{1}},0_{b_{2}},\ldots,0_{b_{s-1}},Y_{c}^{(u,s)},X_{r_{1}}^{(s)},X_{r_{2}}^{(s)},\ldots,X_{r_{n-s}}^{(s)})>t-x\mid
X_{s:n}=x\}\\ &=& P\{Y_{c}^{(u,s)}>t-x\mid
X_{s:n}=x\}\sum_{k=1}^{n-s}p_{k}^{c,(b_{1},\ldots,b_{s-1})}P\{X_{k:n-s}^{(s)}>t-x\mid
X_{s:n}=x\}\\&&+p_{n-s+1}^{c,(b_{1},\ldots,b_{s-1})}
P\{Y_{c}^{(u,s)}>t-x\mid X_{s:n}=x\}.\end{eqnarray*} Thus, the result
follows from Theorem \ref{th1}.$\hfill\Box$\\
Below we consider an example to demonstrate the result of the
theorems.
\begin{Example}\label{aaa}\normalfont
Let us consider the following coherent system having lifetime
$$T = \min(X_{1}, \max(X_{2}, X_{3}))$$ formed by three independent components $C_{1}$, $C_{2}$, $C_{3}$ with lifetimes $X_{1}$, $X_{2}$, $X_{3}$, respectively and a standby component $Y$
which starts to work in cold state and switches over to warm state
after a pre-specified time $u\geq 0$. The signature of this system
is $\textbf{p}=(\frac{1}{3},\frac{2}{3},0)$, and
$k_{\phi}=z_{\phi}=1$ so that $s=1$ or $2$. It is to be noted that,
for $s=1$, $B_{1,1}=\phi$ (as there is no previously failed
component) and $P\{V_{1}=1\}=1$, $P\{V_{1}=2\}=P\{V_{1}=3\}=0$. Now,
from Proposition \ref{prop1}, we have
\begin{eqnarray*}
&&P\{\phi_{1}(Y_{1}^{(u,1)},X_{2}^{(1)},X_{3}^{(1)})>t-x\mid
X_{1:3}=x\}\\
&=&\frac{\bar{G}(\gamma(x-u))}{\bar{G}(\omega(x-u))(\bar{F}(x))^2} \iiint\limits_{\min(y+u,\max(x_{2},x_{3}))>t-x}
g(\omega(x-u)+y+u)f(x_{2}+x)f(x_{3}+x)dx_{2}dx_{3}dy\\&=&\frac{\bar{G}(\gamma(x-u))}{\bar{G}(\omega(x-u))(\bar{F}(x))^2}
\bar{G}(\omega(x-u)+t-x)
\big[(\bar{F}(t))^2+2\bar{F}(t)(F(t)-F(x))\big].\end{eqnarray*} For
$s=2$, previously failed component is either $C_{2}$ or $C_{3}$. In
this case $P\{V_{2}=1\}=\frac{1}{2}$,
$P\{V_{2}=2\}=P\{V_{2}=3\}=\frac{1}{4}$ and
$P\{B_{2,2}=3\}=P\{B_{2,3}=2\}=1$. Now, from Proposition \ref{prop1}, we have
\begin{eqnarray}\nonumber
&&P\{\phi_{2}(0_{2},Y_{1}^{(u,2)},X_{3}^{(2)})>t-x\mid X_{2:3}=x\}
\\ \nonumber
&=&\frac{\bar{G}(\gamma(x-u))}{\bar{G}(\omega(x-u))\bar{F}(x)}
\iint\limits_{\min(y+u,,x_{3})>t-x}
g(\omega(x-u)+y+u)f(x_{3}+x)dx_{3}dy\\\label{eqs2}&=&\frac{\bar{G}(\gamma(x-u))}{\bar{G}(\omega(x-u))\bar{F}(x)}
\bar{G}(\omega(x-u)+t-x) \bar{F}(t).\end{eqnarray} Because $X_{1}$,
$X_{2}$ and $X_{3}$ are independent with common c.d.f. $F$, the
expression for $P\{\phi_{2}(0_{2},Y_{3}^{(u,2)},X_{1}^{(2)})>t-x\mid
X_{2:3}=x\}$, $P\{\phi_{2}(0_{3},Y_{1}^{(u,2)},X_{2}^{(2)})>t-x\mid
X_{2:3}=x\}$, and
$P\{\phi_{2}(0_{3},Y_{2}^{(u,2)},X_{1}^{(2)})>t-x\mid X_{2:3}=x\}$
will be the same as given in (\ref{eqs2}). Thus, from Theorem
\ref{th1}, we have {\footnotesize
\begin{eqnarray*}P\{T^{gs}>t\}&=&\frac{1}{3}P\{X_{1:3}>t\}+\frac{2}{3}P\{X_{2:3}>t\}+\\&&
\frac{1}{3}\int_{u}^t
\frac{\bar{G}(\gamma(x-u))}{\bar{G}(\omega(x-u))(\bar{F}(x))^2}
\bar{G}(\omega(x-u)+t-x)
\big[(\bar{F}(t))^2+2\bar{F}(t)(F(t)-F(x))\big]dF_{1:3}(x)\\&&
+\frac{2}{3}\int_{u}^t
\frac{\bar{G}(\gamma(x-u))}{\bar{G}(\omega(x-u))\bar{F}(x)}
\bar{G}(\omega(x-u)+t-x) \bar{F}(t)dF_{2:3}(x)\\&=&
\frac{1}{3}P\{X_{1:3}>t\}+\frac{2}{3}P\{X_{2:3}>t\}+\\&& \int_{u}^t
\frac{\bar{G}(\gamma(x-u))}{\bar{G}(\omega(x-u))(\bar{F}(x))^2}
\bar{G}(\omega(x-u)+t-x)
\big[(\bar{F}(t))^2+2\bar{F}(t)(F(t)-F(x))\big]F(x)\bar{F}(x)f(x)
dx\\&&+4\int_{u}^t
\frac{\bar{G}(\gamma(x-u))}{\bar{G}(\omega(x-u))\bar{F}(x)}
\bar{G}(\omega(x-u)+t-x) \bar{F}(t)\bar{F}^2(x)f(x)
dx.\end{eqnarray*}} In Fig. 1, we plot the reliability function of
the above coherent system with and without a standby unit when
$F(t)=G(t)=1-e^{-2t}, t>0$ and $\omega(t)=\gamma(t)=t/2$.
\end{Example} \begin{figure}\begin{center}
  \includegraphics[width=12.5cm]{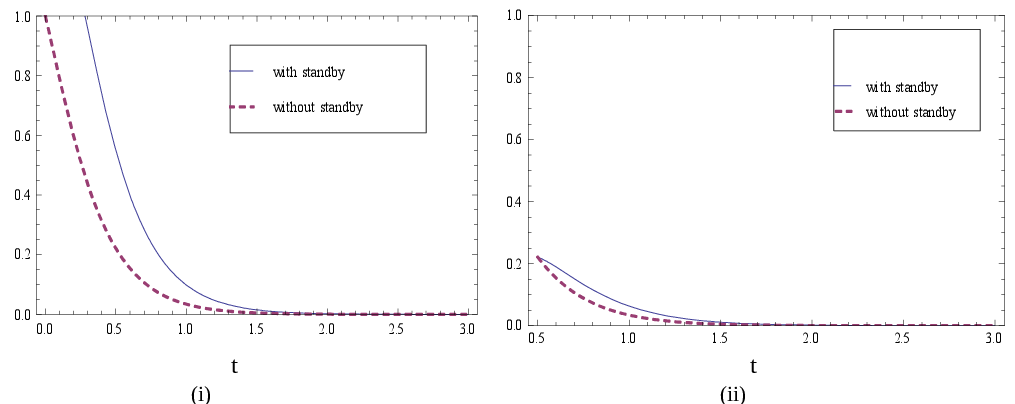}\\\caption{Reliability function of the system with and without standby in case of (i) $u=0$ and (ii) $u=0.5.$}
 \end{center}
\end{figure}
\subsection{Reliability of a coherent system
with imperfect switching} Consider the coherent system as discussed
in Section \ref{sec2.1} with all the assumptions remaining the same
except the switching case. By imperfect switching we mean that the
standby component may fail at the time of switch over from one state
to another with certain positive probability. In our discussion, a
standby component remains in cold state during the period $[0,u)$
and switches over to the warm state at time $u$, when it is as good
as new. So, we assume that at the time of switch over from cold
state to warm state, the standby component does not fail whereas, at
the time of switch over from warm state to active state it does not
fail with pre-assigned probability $p\in[0,1]$. If a standby
component does not fail during switch over, this is known as perfect
switching. We also assume that the change over from one state to
another is instantaneous. Then the reliability of the system is
given by the following theorem. The proof is similar to that of
Theorem \ref{th1}, and hence omitted.
\begin{Theorem}\label{th3}  Let
$\textbf{p}$ be the signature vector of a coherent system as
mentioned above. If the probability of perfect switching from warm state
to active state is $p$, then, for $t>u$,
\begin{eqnarray*}&& P\{T^{gs}>t\}=\sum_{s=k_{\phi}}^{z_{\phi}+1}\big(p_{s}P\{X_{s:n}>t\}+p_{s}\sum_{c=1}^nP\{V_{s}=c\}\sum_{1\leq b_{1}<\ldots<b_{s-1}\leq n}P\{B_{s,c}=(b_{1},\ldots,b_{s-1})\}\\&&
 \times p \int_{u}^t P\{\phi_{s}(0_{b_{1}},0_{b_{2}},\ldots,0_{b_{s-1}},Y_{c}^{(u,s)},X_{r_{1}}^{(s)},X_{r_{2}}^{(s)},\ldots,X_{r_{n-s}}^{(s)})>t-x\mid X_{s:n}=x\}dF_{s:n}(x)\big).\end{eqnarray*}
\end{Theorem}
\begin{Corollary}
    Putting $p=1$ in Theorem \ref{th3} we get Theorem \ref{th1}.
\end{Corollary}
\begin{Remark}\normalfont
    Considering the same coherent system as in Example \ref{aaa}, with probability of perfect switching from warm state to active
state as $p$, we have
\begin{eqnarray*}
P\{T^{gs}>t\}&=&\frac{1}{3}P\{X_{1:3}>t\}+\frac{2}{3}P\{X_{2:3}>t\}+
p\times \\&& \left[\frac{1}{3}\int_{u}^t
\frac{\bar{G}(\gamma(x-u))}{\bar{G}(\omega(x-u))(\bar{F}(x))^2}
\bar{G}(\omega(x-u)+t-x)\right.\\&&\left.
\big[(\bar{F}(t))^2+2\bar{F}(t)(F(t)-F(x))\big]dF_{1:3}(x)\right.\\&&
\left. +\frac{2}{3}\int_{u}^t
\frac{\bar{G}(\gamma(x-u))}{\bar{G}(\omega(x-u))\bar{F}(x)}
\bar{G}(\omega(x-u)+t-x) \bar{F}(t)dF_{2:3}(x)\right]\end{eqnarray*}
\end{Remark}
\subsection{Reliability of a coherent system with Random warm up period}
Consider the coherent system as discussed in Section \ref{sec2.1}
with all the assumptions remaining the same except that switching
from cold state to warm state is not instantaneous and a random warm
up period may be required for state change. Further we consider that
warm up starts at a fixed $u_{1}\in [0,u]$ so that the distribution
function $H$ of the random warm up time has the support $[u_{1},u]$.
We assume that the warm up time is independent of the life of the
standby component. Then the reliability of the system is given by
the following theorem. The proof is similar to that of Theorem
\ref{th1}.
\begin{Theorem}\label{th4}
    Let $\textbf{p}$ be the signature vector of a coherent system as
mentioned above. Then, for $t>u$,
\begin{eqnarray*}\nonumber P\{T^{gs}>t\}&=&\sum_{s=k_{\phi}}^{z_{\phi}+1}\big(p_{s}P\{X_{s:n}>t\}+p_{s}\sum_{c=1}^nP\{V_{s}=c\}\sum_{1\leq b_{1}<\ldots<b_{s-1}\leq n}P\{B_{s,c}=(b_{1},\ldots,b_{s-1})\}
 \times\\&& \int_{u}^t \int_{u_{1}}^{u}P\{\phi_{s}(0_{b_{1}},0_{b_{2}},\ldots,0_{b_{s-1}},Y_{c}^{(r,u_{1},u,s)},X_{r_{1}}^{(s)},X_{r_{2}}^{(s)},\ldots,X_{r_{n-s}}^{(s)})>t-x\mid X_{s:n}=x\}\\&& dH(r) dF_{s:n}(x)\big),\end{eqnarray*}
where $Y_{c}^{(r,u_{1},u,s)}$ is same as $Y_{c}^{(u,s)}$ with warm
up time $r\in[u_{1},u]$.
\end{Theorem}
\begin{Corollary} If warm up time is a degenerate random variable
degenerate at $u$, then Theorem \ref{th4} gives Theorem \ref{th1}.
\end{Corollary}
\begin{Remark} From Proposition \ref{prop1} it can be seen that
\begin{eqnarray*}
&&P\{\phi_{s}(0_{b_{1}},0_{b_{2}},\ldots,0_{b_{s-1}},Y_{c}^{(r,u_{1},u,s)},X_{r_{1}}^{(s)},X_{r_{2}}^{(s)},\ldots,X_{r_{n-s}}^{(s)})>t-x\mid
X_{s:n}=x\}\\
&&=\frac{\bar{G}(\gamma(x-u_{1}-r))}{\bar{G}(\omega(x-u_{1}-r))\bar{F}^{n-s}(x)}
\idotsint\limits_{\phi_{s}(0_{b_{1}},0_{b_{2}},\ldots,0_{b_{s-1}},y+u_{1}+r,x_{r_{1}},x_{r_{2}},\ldots,x_{r_{n-s}})>t-x}
\hspace{-1.5cm}g(\omega(x-u_{1}-r)+y+u_{1}+r)\\&&\prod_{m=1}^{n-s}
f(x_{r_{m}}+x) dx_{r_{1}}dx_{r_{2}}\ldots dx_{r_{n-s}}dy
\end{eqnarray*}
\end{Remark}
    \begin{Remark}\normalfont
Consider the same coherent system as in Example \ref{aaa}. Now if
switching from cold state to warm state requires warm up time and
$H$ is the c.d.f. of warming up time, then we have
\begin{eqnarray*}
P\{T^{gs}>t\}&=&\frac{1}{3}P\{X_{1:3}>t\}+\frac{2}{3}P\{X_{2:3}>t\}+\\&&
\frac{1}{3}\int_{u}^t \int_{u_{1}}^{u}
\frac{\bar{G}(\gamma(x-u_{1}-r))}{\bar{G}(\omega(x-u_{1}-r))(\bar{F}(x))^2}
\bar{G}(\omega(x-u_{1}-r)+t-x)\\&&
\big[(\bar{F}(t))^2+2\bar{F}(t)(F(t)-F(x))\big]
dH(r)~dF_{1:3}(x) +\\&&\frac{2}{3}\int_{u}^t \int_{u_{1}}^{u}
\frac{\bar{G}(\gamma(x-u_{1}-r))}{\bar{G}(\omega(x-u_{1}-r))\bar{F}(x)}
\bar{G}(\omega(x-u_{1}-r)+t-x) \bar{F}(t)
dH(r)~dF_{2:3}(x).\end{eqnarray*}
\end{Remark}
\section{Conclusion}\label{sec4} Although the coherent system with a cold or a
hot standby component has been studied in the literature, to the
best of our knowledge, the coherent system with general standby
component has not been investigated. In this paper, we study the
reliability of a coherent system equipped with a general standby
component. From practical experience, sometimes we get a system,
which tells that at least for some initial period of time, say $u$,
the system will never fail. Keeping this in mind, we allow the
standby component to be in cold state initially for a period
$[0,u]$, where $u$ is fixed and it switches over to the warm state
at time $u$. Because at time $u$, the standby component is as good
as new, it is logical to assume that at the time of switch
over from cold state to warm state it does not fail. We also
consider the case where the standby component may fail at the time
of switch over from warm state to active state (at the failure of a
component which makes the system to fail). This is called imperfect
switch over. In both the cases we get the explicit expression for
the reliability of the system. Since the results in this paper are
more general, the mathematical expressions look complicated. We have
given one example to demonstrate how one can calculate the
reliability of the system in practice. This will surely help people
to use the results in practice.\par Further, in some cases, the
standby component may require some warm up period before switching
over from the cold state to the warm state so that once the system
fails due to failure of a particular component the standby component
may replace the failed component in order to keep the system
functioning. This warm up period may not be known in general. Also
the warming up time will vary depending on the nature of the standby
components. To handle this case, we assume that the warm up time is
a random variable having distribution function $H$ with support
$[u_{1},u]$, $0\leq u_{1}\leq u$.\par It is to be noted that from
our general results, the results related to a $k$-out-of-$n$ system
and the results related to cold standby and hot standby available in
the literature are obtained as particular cases.
\subsection*{Acknowledgements:} The authors are thankful to the reviewers and Prof. D. Paindaveine, the Co-Editor-in-Chief for their valuable comments and
suggestions leading to an improved version of the paper. The support
received from IISER Kolkata to carry out this research work is
gratefully acknowledged by Pradip Kundu. Nil Kamal Hazra sincerely
acknowledges the finencial support from the University of the Free
State, South Africa. The financial support from NBHM, Govt. of India
(vide Ref. No. 2/48(25)/2014/NBHM(R.P.)/R\&D II/1393 dt. Feb. 3,
2015) is duly acknowledged by Asok K. Nanda.

\end{document}